\documentclass[11pt,twoside]{article}
\usepackage{asp2010}

\resetcounters

\aspvoltitle{
Progress in Physics of the Sun and Stars: A New Era in Helio- and Asteroseismology
}
\aspvolauthor{H. Shibahashi and A. E. Lynas-Gray, eds.}

\begin{document}

\title{Frequency Regularities in $\mbox{\boldmath$\delta$}$\,Scuti Stars}
\author{M. \textsc{Papar\'o} and J. M. \textsc{Benk\H{o}}}
\affil{Konkoly Observatory, MTA CSFK, Konkoly-Thege \'ut 15-17, 1121 Budapest, Hungary}

\begin{abstract}
Space missions have produced an incredibly large 
database on pulsating stars. The light curves via the frequency content 
contain a detailed description of each star. The critical point is the 
identification of modes, especially in the non-asymptotic regime. The best 
derived parameters from the frequency content of a pulsating star light curve are the 
frequency differences and ratios. This presentation focuses on the 
potential of period ratios in mode identification. 
\end{abstract}

\section{Introduction}

Mode identification is one of the most critical objectives of asteroseismology. 
In the asymptotic regime (solar type oscillation and white dwarfs) the regularities (patterns) 
help to identify modes and allow us to reach a level of real asteroseismology. 
However, over most of the Hertzsprung Russell Diagram the pulsation is outside the asymptotic regime. 
In particular, $\delta$ Scuti stars exhibit a large discrepancy between the number of observed
and theoretically predicted modes, which points to the  operation of a mode selection mechanism. 
Although we recognized that independent identification of modes is impossible in a large number of stars, 
there always has been a hope that longer and continuous observation would reveal the predicted 
but missing frequencies, or that we will find some kind of regularity amongst the larger number and larger amplitude modes. 
With the Wide Field Infrared Explorer (WIRE),
Microvariability and Oscillations of Stars (MOST), 
Convection Rotation and Planetary Transits (CoRoT) 
and {\it Kepler} space missions our dream has become a reality 
with long and almost continuous data sets now available. 

We have started to perform a systematic check as to how we can use the increased number and most precisely derived observables,
the frequency differences and period ratios. 
The primary test case was CoRoT\,102749568, a $\delta$ Scuti star where 
more than one radial period ratio was immediately noticed. 

\section{Period Spacings of CoRoT\,102749568}

The period spacings of the independent modes in CoRoT\,102749568 have been 
successfully investigated and presented in \citet{pa2013}. Although 
this $\delta$\,Scuti star is also not in the asymptotic regime, the high 
level regularity amongst the frequencies allowed us to derive
 the large separation observationally which assumes a frequent appearance of a
characteristic spacing between the consecutive radial orders. 

\begin{figure}
\centering
  \includegraphics[angle=00,width=\textwidth]
  {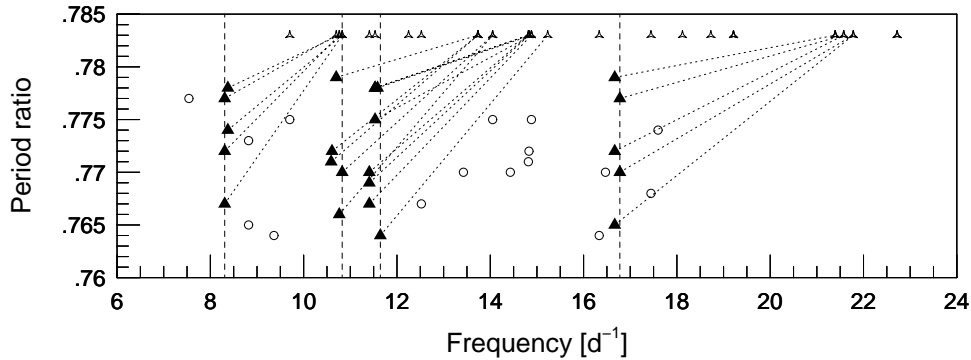}
  \caption{The distribution of the near period ratios in 
CoRoT\,102749568. The vertical dashed lines mark the frequencies where the 
period ratios are distributed as straight lines (at the lower value for 
each pair). Dotted lines give the connection to the higher frequency 
member of each pair (marked at a constant period ratio value). 
The other near radial period ratios are given by open circles.}
\label{near}
\end{figure}

The investigation revealed that not only the radial modes but also nonradial 
modes with different $\ell$ values contribute to the distinguished peak in the spacing.
According to the radial period ratio sequences, modes with three different 
$\ell$ values were localized in the excited modes including five consecutive 
radial orders of the radial mode. 
Not only was the large separation observationally determined and confirmed 
by modelling but the twelve highest amplitude modes were identified using only frequency regularities. 
The presence of the large separation between the modes suggest that these 
modes appear in the region of the near radial period ratio interval. 
Figure~\ref{near} shows a non-homogeneous distribution; the closely spaced 
period ratios create {\it vertical straight lines} in the frequency versus
period ratio diagram due to the grouping of frequencies at one 
value of the frequency pairs. The region of 0.76 -- 0.78 was chosen 
around the canonical value of the period ratio of the radial fundamental and first overtone.
Knowing the identification of modes in CoRoT\,102749568, we can state 
that the vertical lines are around the non-radial modes with $\ell = 1$ and $2$.

\section{Near Period Ratios for Other $\mbox{\boldmath$\delta$}$\,Scuti Stars}

A large sample of $\delta$\,Scuti stars (using both ground-based and 
space data) were checked for generality in the appearance of radial 
period ratios or sequences and how we can use them, if we can, for mode identification.
The stars chosen according to the type of observation, the 
evolutionary stage and rotation are given in Table~\ref{tabl:sample}.
 
\begin{table}[t]
 \caption{A sample of $\delta$\,Scuti stars was investigated to see 
the distribution of period ratios in the near radial period ratio interval.} 
\medskip
 \label{tabl:sample}
 \footnotesize
 \begin{center}
 \begin{tabular}{ccccrrcc}
  \hline\hline
\noalign{\smallskip}
Name &  Type of & Evol. stage & $T_{\mathrm{eff}}$ & $\log g/\log (L/L_\odot)$ & $v\sin i$ & Remarks & Ref.\footnotemark[1]\\
     &  obs.            &             &  (K) &     &       (km\,s$^{-1}$) & &\\
\noalign{\smallskip}
  \hline
\noalign{\smallskip}
44~Tau  &   ground  &  post-MS & 6900   &  3.6 & 3 &  & Br08\\
4~CVn  &   ground  & hi. evolv. & 6900   &  3.5 &  -- &  & Br99\\
XX~Pyx &   ground  &  early MS &  8300  &  4.25  & 52  &  & H00\\
BI~CMi &  ground & cool border & 6925 & 3.69 & 76  &  & Br02\\
FG~Vir &  ground & -- & 7500 & 3.95 & 21.3  & & Br05\\
101155310 & CoRoT & -- & 7300 & 3.75 & 23 & HADS\footnotemark[2]  & P11\\
9408694 & Kepler & -- & 7300 & 3.5 & 100 & HADS\footnotemark[2]  & Ba12b\\
$\epsilon$~Cep &  WIRE &  -- & 7340 & 3.9 & 90 & reg. spac. & Br07 \\
$\alpha$~Oph &  MOST & -- & 8336 & 1.496 & 239 & binary & M10\\
V589~Mon & MOST & pre-MS & 6800 & 1.58 & 60  & &Z11\\
BS~Cnc &  MOST & cluster & 7600 & 4.2 & 135 &   &Br12\\
10661783 &  Kepler & evolved & 8000 & -- & -- & ecl. bin. &S11\\
4840675 &   Kepler  & sl. evolv. & 7400   &  1.1 &  220  & rot. mod. & Ba12a\\
9700322 &   Kepler &  -- &  6700  &  3.7  & 19  & rot. mod. & Br11\\
HD\,174936 & CoRoT & --  & 8000 & 4.08 & 170  & reg. spac. &G09\\
HD\,50844 & CoRoT & evolved & 7500 & 3.6 & 58 & $\lambda$~Bootis & P09\\
\noalign{\smallskip}
  \hline \\
 \end{tabular}
\end{center}
\noindent{
{\scriptsize 1} -- References: Br08: \cite{BrLenz2008},  
Br99: \cite{BrHand99}, 
H00: \cite{Hand2000}, 
Br02: \cite{BrGar2002}, 
Br05: \cite{BrLenz2005}, 
P11: \cite{PoRai2011}, 
Ba12b: \cite{ba2012b}, 
Br07: \cite{bu2007}, 
M10: \cite{Mon2010}, 
Z11: \cite{Zw2011}, 
B12: \cite{BrHar2012}, 
S11: \cite{South2011}, 
Ba12a: \cite{ba2012},
Br11: \cite{br2011}, 
G09: \cite{GarMoy2009}, 
P09: \cite{po2009}}\\ \\
\noindent{
{\scriptsize 2} -- High Amplitude Delta Scuti (HADS)}
\end{table}

\begin{figure}[h!]
\centering
  \includegraphics[angle=00,width=\textwidth]
  {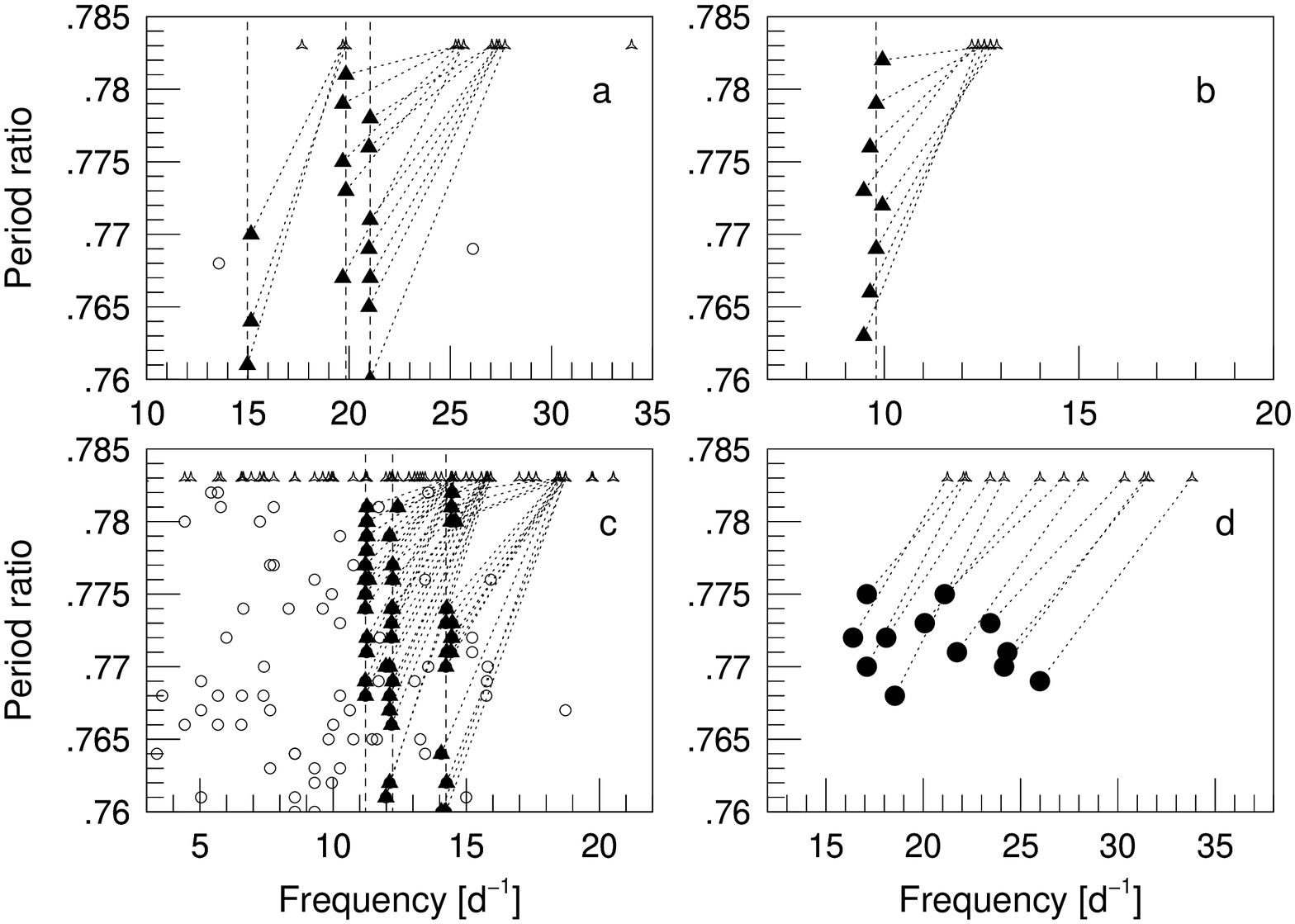}
  \caption{The distribution of the near period ratios in selected $\delta$\,Scuti stars: 
$\epsilon$~Cep (a), KIC\,9700322 (b), HD\,50844 (CoRoT, c) and KIC\,4840675 (d). The 
appearance of the period ratios in the near radial period ratio range shows that pairs of 
consecutive radial orders exist. The vertical lines show grouping of modes around one or both 
radial orders.}
  \label{sampl}
\end{figure}

In Fig.\,\ref{sampl} we present four typical cases though we have some general conclusions. 
The near radial period ratio region is highly populated and the period ratios are not homogeneously or randomly distributed. 
If frequencies near the frequency pairs separated by the large separation are grouped either at the 
lower or at the higher frequency value of the pairs, then straight lines are created from the 
closely spaced values in the near radial period ratio range.
The range of grouping is frequency dependent ($\Delta f$ = 0.3\,--\,0.6\,d$^{-1}$ for CoRoT\,102749568).
When we do not know the rich frequency content (most of the ground-based samples) then we do not have 
a structure. The opposite case appears when the frequency content is very rich (some of the 
space data) with many low amplitude modes, then the distribution is very dense and any kind of 
regularity is hardly noticed.
In this case we investigated only the frequency content with the highest amplitude 
(HD\,50844, \citealt{po2009}). Our sample shows that in fast rotating 
stars there are no 
regular 
structures in the near radial period ratio range (KIC\,4840675, \citealt{ba2012}). In 
$\epsilon$~Cep \citep{bu2007} straight lines appear around both the radial ($\ell = 0$) and 
nonradial ($\ell = 1, 2$) modes. 
A grouping around radial modes may appear if non-radial trapped modes are close to the 
radial modes. KIC\,9700322 \citep{br2011} shows a simple case of a rotationally modulated 
star. The rotational splittings also create a straight line in the near radial period ratio range.

\section{Test for Period Ratios}
\begin{figure}[!ht]
\centering
  \includegraphics[angle=00,width=\textwidth]{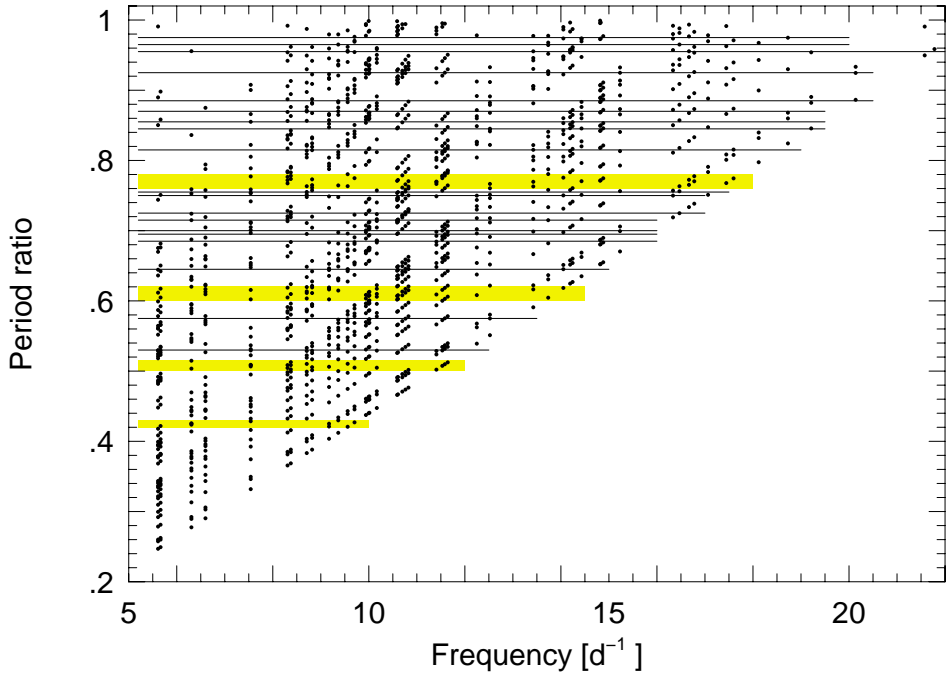}
  \caption{
The whole range of period ratios for CoRoT\,102749568. The marked horizontal segments are used for mode identification. 
The segments marked in yellow give the period ratio of the radial overtones to the radial fundamental mode. 
To avoid the overlapping, the other segments are represented only by lines.
The lines give the mean value of the segments for the other combination of the radial orders of the different $\ell$ values ($\ell = 0$, 1 and 2). 
Table\,\ref{tabl:seg} gives the exact period ratio range of the selected segments.
}
  \label{vertical}
\end{figure}

The previously investigated near radial period region is important to get the grouping of frequencies around modes separated by the large separation. However, a wider range of period ratios (which has never been used before) can be derived from the frequency content as we show in Fig.\,\ref{vertical}. These values contain information not only for the radial modes, or modes separated by the large separation, but also for all the modes excited in the star.

The period ratios (equivalent to frequency ratios) were calculated for the 52 independent modes of 
CoRoT\,102749568 such as
$f_1/f_2$, $f_1/f_3$, $f_1/f_4$, $\dots$, $f_2/f_3$, $f_2/f_4$,
$\dots$, $f_{51}/f_{52}$.
Here indices indicate the position of the frequency in the order:
$f_1$ is the lowest frequency and $f_{52}$ is the highest one.
When we plot the period ratio versus the frequency (in the counters)
we get Fig.~\ref{vertical}. Some structures of the figure
can simply be explained from its construction. The ``vertical structures''
show the period ratios connected to the frequencies in the counters, the slope of the
``inclined lines'' indicates the 
frequencies in the denominators. Finally, the distribution of the period ratios in 
both lines reflects the frequency distribution itself. Closely spaced period ratios appear 
everywhere but their actual place depends on the grouping of the frequencies.

\begin{figure}[!ht]
\centering
  \includegraphics[angle=00,width=\textwidth]
  {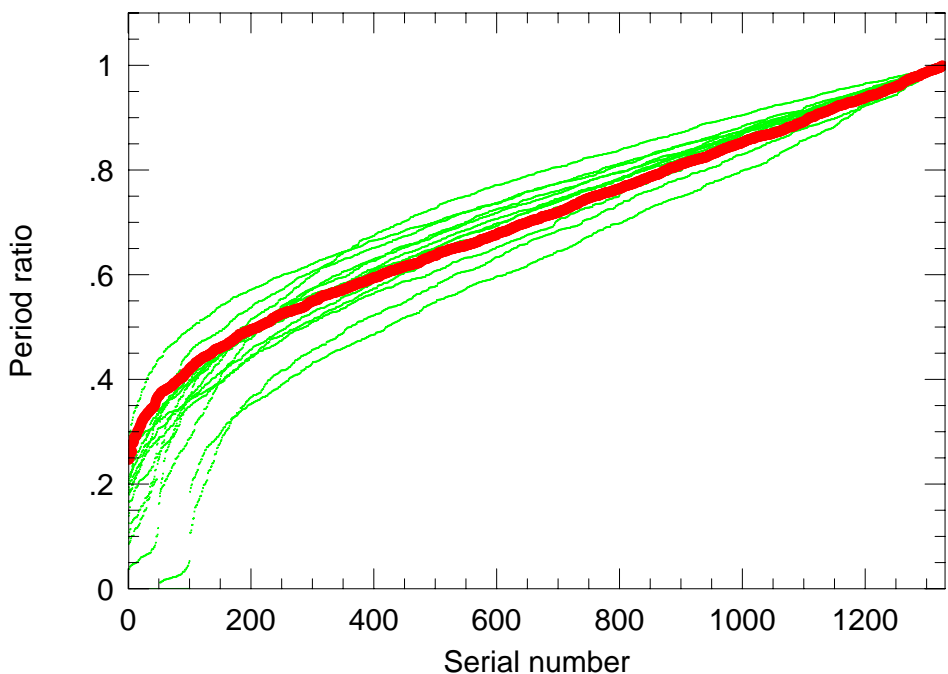}
  \caption{Comparison of the random distribution to CoRoT\,102749568. The thick (red) curve 
gives the observed distribution of ratios. The thin (green) curves represent some 
distributions of ratios from random samples.}
  \label{distr}
\end{figure}

The important question is whether the distribution
of ratios really reflects the structure of the pulsation or not.
To check this we carried out a similar process for a randomly generated 
frequency content with the same averages and standard deviations as the observed set. 
We ordered the total number of ratios
by their values and plotted them successively; namely the index 1 belongs to the lowest 
ratio while the highest ratio has the index 1326 (see Fig.\,\ref{distr}).

We did not detect any serious agglomerations in the ratios since we did not see
a kind of staircase function.
The distribution was rather smooth in both cases and seemed to be similar on the face of it.
To verify the similarity in a mathematical way we performed a
Monte Carlo simulation: 1\,000 random frequency sets were prepared
and processed in the same manner as the observed one.
Then a two-sample Kolmogorov-Smirnov (KS) test \citep{press01}
was applied for each pair formed from a random sample and the original
data set.
The test rejected the null hypothesis with a probability of  $0.92\pm0.17$.
That is, the period ratio distribution of CoRoT\,102749568 definitely has a 
different structure than the random one.

\section{Mode Identification According to Period Ratios in CoRoT\,102749568}
 
\begin{table}[!b]
 \caption{Selected horizontal segments.}
 \label{tabl:seg}
 \medskip
 \centering
 \begin{tabular}{rclrclrcl}
  \hline\hline
\noalign{\smallskip}
$\ell, n$ & $\ell, n$ & $\;\;\;\;\;$range & $\ell, n $ & $\ell, n$ & $\;\;\;\;\;$range & $\ell, n$ & $\ell, n$ & $\;\;\;\;\;$range \\
\noalign{\smallskip}
  \hline
\noalign{\smallskip}
0, 0 & 0, 1 & 0.76 -- 0.78  & 0, 2 & 2, 2 & 0.72 -- 0.73$\;\:\:$  & 0, 0 & 2, 1 & 0.525 -- 0.535 \\
0, 0 & 0, 2 & 0.60 -- 0.62  & 1, 0 & 2, 0 & 0.71 -- 0.72  & 0, 1 & 2, 2 & 0.57$\;\:$ -- 0.58 \\
0, 0 & 0, 3 & 0.50 -- 0.515$\:$ & 1, 1 & 2, 1 & 0.75 -- 0.76  & 1, 1 & 2, 0 & 0.925 -- 0.935 \\
0, 0 & 0, 4 & 0.42 -- 0.43  & 0, 0 & 1, 1 & 0.69 -- 0.71  & 1, 2 & 2, 1 & 0.96$\;\:$ -- 0.97 \\
2, 0 & 2, 1 & 0.81 -- 0.82  & 2, 0 & 1, 2 & 0.84 -- 0.85  & 1, 3 & 2, 2 & 0.97$\;\:$ -- 0.98 \\
2, 0 & 2, 2 & 0.69 -- 0.70  & 0, 2 & 1, 3 & 0.74 -- 0.75  & 2, 0 & 0, 2 & 0.95$\;\:$ -- 0.96 \\
0, 0 & 2, 0 & 0.64 -- 0.65  & 1, 0 & 0, 1 & 0.85 -- 0.86  & 2, 1 & 1, 3 & 0.865 -- 0.875 \\
0, 1 & 2, 1 & 0.68 -- 0.69  & 1, 1 & 0, 2 & 0.88 -- 0.89  &      &      &        \\
\noalign{\smallskip}
  \hline
 \end{tabular}
\end{table}

All information about the star, the frequency content, the grouping of frequencies, the spacing 
of the consecutive radial orders of a given $\ell$ and the spacings of the modes with different $\ell$ 
values are given in the whole range of the period ratios (Fig.\,\ref{vertical}). 
The key to mode identification is knowing which horizontal segments have to be used. 
Knowing the identification of twelve modes in CoRoT\,102749568, we selected some critical 
horizontal segments which reflect the large separation and the characteristic spacings 
between the modes with different $\ell$ values. The $\ell$ and $n$ for a given period ratio and 
the actual ranges (marked on Fig.\,\ref{vertical}) are given in Table~\ref{tabl:seg}. 
Each segment determines two subsets of frequencies that are connected to each other by the 
given period ratio. Due to the grouping, some frequencies are missing.
For mode identification we used the fact that the missing frequencies may not have the 
same quantum numbers as we used for the segment. 
As a consequence of the exclusions, we got a mode identification of the twelve modes 
with rather high probability. For three modes we got a unique 
identification ($\ell=0,\ n=1$; $\ell=2, \ n=1$ 
and $\ell=1, \ n=0$). For 3 modes we got 83\% probability and 75\% for another 3 modes. 
Only three modes had a weak identification (60\%). The real identification of 
only one mode ($\ell=0, \ n=4$) was excluded in our process which means that we did not 
select segments sensitive enough for this mode.
    
A known identification of modes was used in our test case to select the segments. 
In real cases we do not have any a priori knowledge. However, if the critical range of 
segments are calibrated by the quantum numbers of the modes excited in models in general, 
we could use the process to get a tentative identification of modes in the large number
of stars observed by space instruments.

\acknowledgements M. Papar\'o  and J. M. Benk\H o acknowledge the support of the 
European Space Agency (ESA) Plan for European Cooperating States (PECS) project
No. 4000103541/11/NL/KML.

\bibliography{paparo}

\end{document}